\documentclass[12pt]{article}
\usepackage{amsmath,amssymb,amsfonts}
\usepackage[paper=letterpaper,margin=1.0in]{geometry}
\usepackage{graphicx}

\newcommand{\be}{\begin{equation}}
\newcommand{\ee}{\end{equation}}
\newcommand{\bea}{\begin{eqnarray}}
\newcommand{\eea}{\end{eqnarray}}
\newcommand{\ba}{\begin{array}}
\newcommand{\ea}{\end{array}}
\newcommand{\ben}{\begin{enumerate}}
\newcommand{\een}{\end{enumerate}}
\newcommand{\bi}{\begin{itemize}}
\newcommand{\ei}{\end{itemize}}
\newcommand{\bc}{\begin{center}}
\newcommand{\ec}{\end{center}}
\newcommand{\bfig}{\begin{figure}}
\newcommand{\efig}{\end{figure}}
\newcommand{\bq}{\begin{quotation}}
\newcommand{\eq}{\end{quotation}}
\newcommand{\bt}{\begin{table}}
\newcommand{\et}{\end{table}}
\newcommand{\btab}{\begin{tabular}}
\newcommand{\etab}{\end{tabular}}
\newcommand{\bs}{\begin{slide}}
\newcommand{\es}{\end{slide}}

\begin{document}

{\footnotesize
${}$
}

\bc

\vskip 1.0cm
\centerline{\Large \bf Towards a holographic theory of cosmology -- 
threads in a tapestry}
\vskip 0.5cm
\vskip 1.0cm

\renewcommand{\thefootnote}{\fnsymbol{footnote}}

\centerline{{\bf
Y.\ Jack Ng
\footnote{\tt yjng@physics.unc.edu}
}}

\vskip 0.5cm

{\it
Institute of Field Physics, Department of Physics and
Astronomy,\\
University of North Carolina, Chapel Hill, NC 27599, U.S.A.
}

\vspace{3cm}

This Essay received Honorable Mention in the 2013 Essay Competition of 
the Gravity Research Foundation 

\vspace{1cm}

Dedicated to the memory of Hendrik van Dam (1934-2013)
\ec

\vskip 1.0cm

\begin{abstract}

In this Essay we address several fundamental issues in 
cosmology: What is the nature of dark 
energy and dark matter?  Why is the dark sector so
different from ordinary matter?  Why is the effective cosmological 
constant non-zero but so incredibly small? What is the 
reason behind the emergence of a critical acceleration parameter 
of magnitude $10^{-8} cm/sec^2$ in galactic dynamics?
We suggest that the holographic principle is the linchpin in a
unified scheme to understand these various issues.

\end{abstract}

\renewcommand{\thefootnote}{\arabic{footnote}}

\newpage

\section{Introduction: A holographic theory of cosmology}

Some alert readers may have already noticed a resemblance between
the title of this Essay and that of S. Glashow's Nobel Lecture 
\cite{SLG}
``Towards a unified theory -- threads in a tapestry."  This
resemblance is not a coincidence, for like elmentary particle
physics, the study of cosmology is like a patchwork quilt.
But whereas the patchwork quilt has become
a tapestry for the former, the various threads have 
yet to be coherently woven for the latter. 
However now there is reason for optimism: we may 
have found a powerful guiding principle 
behind nature's intricate design, yielding (eventually)
a beautiful tapestry of gravity and matter.  We are referring to
the holographic principle \cite{holography,Susskind}, an important 
by-product of the synthesis of 
quantum mechanics and general relativity, 
according to which,
the maximum amount of information stored in a region of space scales as
the area of its two-dimensional surface, like a hologram.
The holographic principle is arguably
the most important concept in quantum gravity, playing a
role similar to the gauge principle in particle physics.

In this Essay we will apply the holographic principle to address 
a few fundamental issues in gravity and cosmology.
One of the key issues in cosmology is to understand the nature 
of dark energy and dark matter and why the dark sector is so
different from ordinary matter.
Another issue is to explain the twin puzzles of why our universe
is at or very close to its critical density and why
the (effective) cosmological constant is nonzero and so small.
At the (smaller) galactic scale, there are the issues of
the observed flat rotation curves and the emergence of
a critical acceleration parameter separating the regime where
Newtonian dynamics works well from that where it appears to fail.
We liken the resolution of all these issues to finding the right 
threads in a tapestry --- interwoven coherently, with one thread 
logically leading to another.

\section{From spacetime foam to cosmological constant $\Lambda$}

As will be shown shortly, all the aforementioned
issues are linked to the quantum nature of spacetime.  Thus it 
behooves us to start by examining
how foamy spacetime is, or, in other words,
how large the quantum fluctuations of spacetime are.\cite{ng94,Karol}
Let us consider mapping
out the geometry of spacetime for a spherical volume of radius $l$ over 
the
amount of time $2l/c$ it takes light to cross the volume.\cite{llo04}  
One way to do this is to fill the space with clocks, 
exchanging
signals with the other clocks and measuring the signals' times of 
arrival.  
The total number of operations, including the ticks of the clocks and
the measurements of signals, is bounded by the Margolus-Levitin
theorem \cite{mar98}
which stipulates that the rate of operations 
cannot exceed the amount of energy $E$ that is available for the 
operation divided by $\pi \hbar/2$.  This theorem, combined with the
bound on the total mass of the clocks to prevent black hole formation, 
implies that the total number of operations that can occur in this 
spacetime volume is no bigger than
$2 (l/l_P)^2 / \pi$, where $l_P = \sqrt{\hbar G/ c^3}$ is the Planck 
length.  
To maximize spatial resolution,
each clock must tick
only once during the entire time period.  If we regard the operations
as partitioning the spacetime volume into ``cells", then on the 
average 
each cell
occupies a spatial volume no less than $\sim l^3 / (l^2 / 
l_P^2)
= l l_P^2$, yielding an average separation between neighboring
cells no less than $ \sim l^{1/3} l_P^{2/3}$.
\cite{ng08}  This spatial separation can be 
interpreted as the average minimum uncertainty in the
measurement of a distance $l$, that is, $\delta l \gtrsim l^{1/3}
l_P^{2/3}$, \footnote{One way to detect this minute fluctuation
is to look for blurry images of distant quasars in powerful
telescope interferometers. \cite{cfnp}}
in {\it agreement} with the result found in the Wigner-Salecker 
gedanken experiment \footnote
{
In the Wigner-Salecker experiment \cite{sal58,ng94},
a light signal is sent from a clock to a mirror (at a distance 
$l$ away) and back to the clock in a timing experiment 
to measure $l$.   
From the jiggling of
the clock's position alone, the uncertainty principle yields
$(\delta l)^2 \gtrsim \hbar l / mc$, where $m$ is the mass of the 
clock.
On the other hand, the clock must be large enough not to
collapse into a black hole; this requires
$\delta l \gtrsim 4Gm/c^2$. 
We conclude that the fluctuation 
of a distance $l$ scales as $\delta l \gtrsim l^{1/3} l_P^{2/3}$.
\cite{ng94,Karol}}
to measure the fluctuation of a distance $l$.

We make two observations: \cite{Arzano,plb} First, maximal
spatial resolution (corresponding to  $\delta l \sim l^{1/3}
l_P^{2/3}$)
is possible only if the maximum energy density 
$\rho \sim (l l_P)^{-2}$
is available to map the geometry
of the spacetime region, without causing a gravitational collapse.
Secondly,
since, on the average, each cell occupies a spatial volume of $l 
l_P^2$,
a spatial region of size $l$ can contain no more than 
$\sim l^3/(l l_P^2) = (l/l_P)^2$ cells.
Hence, this result for spacetime fluctuations 
corresponds to the case of
maximum number of bits of information $l^2 /l_P^2$
in a spatial region of size $l$, that is
allowed by the holographic principle\cite{holography,Susskind}.

It is straightforward to generalize \cite{Arzano} the above discussion 
for a static 
spacetime region with low spatial curvature to the
case of an expanding
universe by the substitution of $l$ by $H^{-1}$ in the expressions for
energy and entropy densities, where $H$ is the Hubble parameter.  
(Henceforth we adopt $c=1=\hbar$ for convenience unless stated 
otherwise for clarity.)  Thus,
applied to cosmology, the above argument leads to the prediction
that (1) {\it the cosmic energy density has the critical 
value} $\rho \sim (H/l_P)^2$, 
and (2) the universe
of Hubble size $R_H$ contains 
$\sim H R_H^3/ l_P^2 \sim (R_H/l_p)^2$ bits of 
information.  It follows that 
the average energy carried by each particle/bit is
$\rho R_H^3/I \sim R_H^{-1}$.
Such long-wavelength {\it constituents of dark energy give rise to
a more or less uniformly distributed cosmic energy density and
act as a dynamical cosmological constant with the observed 
small but nonzero value}
$\Lambda \sim 3 H^2$. \footnote{Here we will not address the
old cosmological constant problem of why it is not of the
Planck scale.  See Ref. \cite{ng92} for possible solutions.}
\cite{ng08}   Later we will show that these
``particles"/bits have exotic statistical properties.

\section{From $\Lambda$ to MoNDian dark matter}

The dynamical cosmological constant (originated from quantum 
fluctuations of spacetime) can now be
shown to give rise to a critical acceleration
parameter in galactic dynamics.  The argument \cite{HMN} is based on
a simple generalization of E. Verlinde's recent proposal 
of entropic gravity \cite{verlinde,Jacob95}.
Consider a particle with mass $m$ approaching a holographic screen
at temperature $T$.  Using the first law of thermodynamics to 
introduce the concept of entropic force
$
F = T \frac{\Delta S}{\Delta x},
$
and invoking Bekenstein's original arguments \cite{bekenstein}
concerning the entropy $S$ of black holes,
$\Delta S = 2\pi k_B \frac{mc}{\hbar} \Delta x$,
we get $ F = 2\pi k_B \frac{mc}{\hbar} T$.  
In a deSitter space with cosmological constant $\Lambda$, the net 
Unruh-Hawking temperature, \cite{unruh,Davies,hawking} as measured by 
a non-inertial observer with acceleration $a$ 
relative to an inertial observer, is 
$T = \frac{\hbar}{2\pi k_B c} [\sqrt{a^2+a_0^2} - a_0]$, \cite{deser}
where $a_0 \equiv \sqrt{\Lambda / 3}$.  Hence the
entropic force (in deSitter space) is given by 
$F =  m [\sqrt{a^2+a_0^2}-a_0]$.
For $ a \gg a_0$, we have $F/m \approx a$ which gives $a = 
a_N \equiv GM/r^2$, the familiar Newtonian value for the acceleration 
due to the source $M$. But for $a \ll a_0$, 
$F \approx m \frac{a^2}{2\,a_0},$ so
the terminal velocity $v$ of the test mass $m$ in a circular motion 
with radius $r$ should be determined from
\,$ m a^2/(2a_0) = m v^2/r$.  In this small acceleration regime,
the observed flat galactic rotation 
curves ($v$ being independent of $r$) now require
$ a \approx \left( a_N \,a_0^3 \, \right)^{\frac14}$.
But that means
$F \approx m \sqrt{a_N a_0}\,$.
This is the celebrated modified Newtonian dynamics (MoND)
scaling \cite{mond,FandM,interpol}, discovered by Milgrom 
who introduced the critical acceleration 
parameter $a_0$ by hand to phenomenologically explain the flat 
galactic rotation curves.
Lo and behold, {\it we have recovered} MoND {\it with the correct
magnitude for the critical galactic acceleration parameter} $a_0 
\sim 10^{-8} cm/s^2$.
From our perspective, MoND is a {\it classical} phenomenological
consequence of {\it quantum} gravity (with the $\hbar$ dependence
in $T \propto \hbar$ and $S \propto 1/\hbar$ cancelled out). 
\cite{HMN}  As a bonus, we have also recovered 
the observed Tully-Fisher relation ($v^4 \propto M$).

Having generalized Newton's 2nd law, we \cite{HMN}
can now follow the second half of Verlinde's argument \cite{verlinde} 
to generalize Newton's law of gravity
$a= G M /r^2$\, by considering an
imaginary quasi-local (spherical) holographic screen of area $A=4 \pi
r^2$ with temperature $T$.  Invoking the  
equipartition of energy $E= \frac{1}{2} N k_B T$
with $N = Ac^3/(G \hbar)$ being 
the total number of degrees of freedom (bits) on the screen, 
as well as the Unruh
temperature formula and the fact that $E= M_{total} c^2$, we
get 
$2 \pi k_B T 
= G\,M_{total} /r^2$,
where $M_{total} = M + M_d$ represents the \emph{total} mass 
enclosed 
within the volume $V = 4 \pi r^3 / 3$, with
$M_d$ being some unknown mass, i.e., dark
matter.  For $a \gg a_0$, consistency with the Newtonian force law 
$a \approx a_N$ implies $M_d \approx 0$.  But
for $a \ll a_0$, consistency with the condition 
$a \approx \left( a_N \,a_0^3  \right)^{\frac14}$ requires 
$M_d \approx \left(\,\frac{a_0}{a}\,\right)^2\, M
\sim (\sqrt{\Lambda}/G)^{1/2}M^{1/2}r$.  This yields
the dark matter mass density $\rho_d$ profile given by
$
\rho_d(r) \sim M^{1/2}(r_v) (\sqrt{\Lambda}/G)^{1/2} / r^2,
$ 
for an ordinary (visible) matter source of radius $r_v$ with
total mass $M(r_v)$. \footnote{
This result can be compared with the distribution associated with
an isothermal Newtonian sphere in hydrostatic equilibrium (used by 
some dark matter proponents):
$
\rho (r) = \sigma (r^2 + r_0^2)^{-1}.
$
Asymptotically the two expressions agree with
$\sigma$ identified as $\sim M^{1/2}(r_v) (\sqrt{\Lambda}
/G)^{1/2} $. 
}

Thus dark matter indeed exists!
And the MoND {\it force law} derived above, at the galactic scale, 
{\it is simply a manifestation of dark matter}! \cite{turnerkip}
Dark matter of this kind can behave \emph{as if} there is no dark
matter but MoND.  
Therefore, we call it ``MoNDian dark matter".
Intriguingly the dark matter profile we have
obtained relates, at the galactic scale,
dark matter ($M_d$), dark energy ($\Lambda$) and ordinary matter
($M$) to one another.  
Moreover, our theory, unlike the MoND scheme, is
compatible with cosmology, if one properly 
uses a fully relativistic
source (including MoNDian dark matter) at the cluster and cosmic 
scales.\cite{HMN}


\section{Infinite statistics for the dark sector}

Why is the dark sector so different from ordinary matter?  
The reason, as we will show in this section, is 
that {\it the quanta constituting the dark sector obey,
not the familiar Fermi or Bose statistics as for ordinary matter,
but rather an exotic statistics known as the infinite statistics}.
\cite{plb}

First consider the $N \sim (R_H/ l_P)^2$ ``particles" 
constituting dark energy at temperature $T \sim R_H^{-1}$ 
(the average particle energy) in a 
volume $V \sim R_H^3$ that is the whole Hubble volume.
Let us assume that the "particles" obey the familiar Boltzmann 
statistics.  A standard calculation (for the relativistic case) 
yields the
partition function $Z_N = (N!)^{-1} (V / \lambda^3)^N$, where
$\lambda = (\pi)^{2/3} /T$, and the entropy
$S = N [ln (V / N \lambda^3) + 5/2]$.
But now since $V \sim \lambda^3$, the entropy
$S$ becomes nonsensically negative unless $ N \sim
1$ which is equally nonsensical because $N \sim (R_H/l_P)^2 \gg 1$.  
However, if the $N$ inside the log term for $S$ somehow
is absent, then we have a manifestly non-negative
$ S \sim N \sim (R_H/l_P)^2$. 
That is the case if the ``particles" are
distinguishable and nonidentical, for then the Gibbs $1/N!$ 
factor is absent from the partition function $Z_N$.
But the only known consistent statistics in greater than two space
dimensions
without the Gibbs factor is infinite statistics (sometimes called
``quantum Boltzmann statistics") \cite{infinite,infinite2},
as described by the Cuntz algebra
(a curious average of the bosonic and fermionic algebras)
$a_i \, a^{\dagger}_j = \delta_{ij}\,$.  Thus the ``particles" 
constituting dark energy obey infinite statistics. \cite{plb,jejjala}
\footnote{Our result for the
$N \sim (R_H/ l_P)^2$ quanta of dark energy obeying infinite statistics
has received support from Ref. \cite{xiaochen} which shows that the
entropy bound of infinite statistics obeys the area law.}

Next, to show that the quanta of MoNDian dark matter also obey
this exotic statistics, we \cite{PRD}  first reformulate MoND 
via an effective gravitational dielectric medium, 
motivated by the analogy \cite{dielectric} between
Coulomb's law in a dielectric medium and Milgrom's law for MoND.
We start with the nonlinear electrostatics embodied in 
the Born-Infeld theory \cite{bi},
and write the corresponding gravitational Hamiltonian density as
$H_g = b^2 \left(\,\sqrt{1+ D_g^2/b^2}-1\,\right)/(4 \pi)$,
where $D$ stands for the electric displacement vector and $b$ 
is the maximum field strength in the Born-Infeld theory.
With $A_0 \equiv b^2$ and $ \vec{A} \equiv b \, \vec{D_g}$, the 
Hamiltonian density becomes
$H_g = \left(\,\sqrt{A^2+A_0^2}-A_0\,\right)/(4 \pi)$. 
If we invoke energy equipartition
($H_g = \frac{1}{2}\,k_B\, T_{\rm eff}\,$) and
the Unruh temperature formula ($T_{eff} = 
\frac{\hbar}{2\,\pi\,k_B\,c}\, 
a_{\rm eff}\,$), and apply the equivalence principle (in identifying,
at least locally, the local accelerations $\vec{a}$ and $\vec{a}_0$
with the local gravitational fields
$\vec{A}$ and $\vec{A}_0$ respectively), 
then the effective acceleration $a_{\rm{eff}}$
is identified as $a_{\rm eff} \equiv
\sqrt{a^2+a_0^2}-a_0\,$.  But this,
in turn, implies that the Born-Infeld inspired force law takes 
the form (for a given test mass $m$)
$F_{\rm BI} = m\, \left(\,\sqrt{a^2+a_0^2}-a_0\,\right)\,$,
which is precisely the MoNDian force law! 

To be a viable cold dark matter candidate, the quanta of
the MoNDian dark matter must
be much heavier than $k_B\,T_{\rm{eff}}$ since $T_{\rm{eff}}$, 
with its quantum origin (being proportional 
to $\hbar$), is a very low temperature.  Now
recall that the equipartition theorem in general states that
the average of the Hamiltonian is given by
$\langle H \rangle = - \frac{\partial \log{Z(\beta)}}{\partial 
\beta}\,$,
where $\beta^{-1} = k_B T$.  To obtain
$\langle H \rangle = \frac{1}{2} \,k_B\, T$ per degree of 
freedom, even for very low temperature,
we require the partition function $Z$ to be of the Boltzmann form
$Z = \exp(\,-\beta\, H\,)\,$.
But this is precisely the case of infinite statistics. \cite{PRD}




\section{Conclusion: Threads in a tapestry of holography}

In summary, by examining the microscopic
fluctuations of spacetime we have found that
our universe is naturally at or close to its critical density. 
The application of the holographic principle then yields an
effective dynamical cosmological constant of the observed value.
Next we have provided an entropic/holographic interpretation
behind Milgrom's modification of Newton's laws and have uncovered 
a critical galactic acceleration parameter of the correct magnitude 
whose value is intimately related to the dynamical cosmological 
constant. We have also explained how Milgrom's MoND 
can be viewed as a phenomenological manifestation of 
dark matter with a curious mass profile that connects, at the 
galactic scale, the dark matter content to the ordinary matter 
content and dark energy.  In principle this dark matter
mass profile can be checked by observations. \cite{edmonds} 
Last but not least, we have shown that the quanta of the 
dark sector obey infinite statistics; this may explain 
why the dark sector is so different from ordinary matter.

The last result could be profound.  But, if true, it also makes
an analysis of the dark sector considerably more difficult.  The 
reason is that a theory of particles obeying infinite statistics, 
unlike ordinary quantum field theories, is not 
local\footnote{The fields are not local, neither in the 
sense that their observables commute at spacelike separation nor in
the sense that their observables are pointlike functionals of the 
fields.
The expression for the number operator is both nonlocal and non-polynomial
in the field operators, and so is the Hamiltonian.}. \cite{infinite2} 
On the other hand, such a theory of
MoNDian dark matter would be fundamentally quantum gravitational and thus
would give very unusual and distinct yet-to-be explored 
particle phenomenology.

Now if indeed the quanta of the dark sector obey infinite
statistics, then we may wonder whether
quantun gravity is actually the origin of particle statistics  
and whether the underlying statistics is infinite statistics.  Here
is an intriguing thought \cite{PRD}: Is it
possible that ordinary particles that obey Bose or Fermi statistics
are actually some sort of collective degrees of freedom?  (For a
discussion of constructing bosons and fermions out of particles  
obeying infinite statistics, see Ref. \cite{bfquon}
and \cite{shevchenko}.)

Using the holographic principle as our beacon, we
have taken some small yet tightly logical steps
towards a comprehensive understanding of how our universe works
-- from the foaminess of spacetime to the critical cosmic energy 
density, from the dynamical cosmological constant via the holographic
principle to the critical acceleration parameter in local galactic
dynamics, and from dark matter with MoNDian scaling to the dark sector 
obeying infinite statistics.  All these various issues of cosmology
have been found to be inter-related -- like 
inter-connecting patches of a quilt, 
woven together.  Yet this is obviously work in progress.
New threads will have to be added, loose ones to be tightend, and some 
old ones to be overwoven. 
But we are hopeful that the
end product will be a magnificient tapestry.

\vskip 0.5cm

\noindent
{\bf Acknowledgments:}
This Essay is partly based on work done in collaboration with
C.M. Ho and D. Minic, with S. Lloyd, and with M. Arzano and
T. Kephart.  I thank them all.  I would like to dedicate 
this Essay to the memory of my friend and longtime collaborator
H. van Dam who passed away recently,
in witness of my appreciation for him.
This work was supported in part by the US Department of Energy
and the Bahnson Fund of UNC-CH.

\end{document}